\documentclass[aps,prd,reprint,superscriptaddress]{revtex4-1}
\usepackage{amsmath}
\usepackage{amssymb}
\usepackage{graphicx}
\usepackage[caption=false]{subfig}
\usepackage{enumitem}
\usepackage{color}
\usepackage[percent]{overpic}
\usepackage{bm}
\usepackage{rotating}
\usepackage{cancel}
\usepackage{mathtools}
\usepackage{comment}
\usepackage{braket}
\usepackage{hyperref}

\newcommand{\abs}[1]{\left| #1 \right|} 

\newcommand{\ii}{\mathrm{i}}

\newcommand{\px}{+}
\newcommand{\mx}{-}
\newcommand{\pmx}{\pm}
\newcommand{\pz}{e}
\newcommand{\mz}{g}
\newcommand{\scales}{\Upsilon}
\newcommand{\dd}{\mathrm{d}}

\begin{document}

\title{Engineering negative stress-energy densities with quantum energy teleportation}

\author{Nicholas Funai}
\affiliation{Institute for Quantum Computing, University of Waterloo, Waterloo, Ontario, N2L 3G1, Canada}
\affiliation{Department of Applied Mathematics, University of Waterloo, Waterloo, Ontario, N2L 3G1, Canada}
\author{Eduardo Mart\'{i}n-Mart\'{i}nez}
\affiliation{Institute for Quantum Computing, University of Waterloo, Waterloo, Ontario, N2L 3G1, Canada}
\affiliation{Department of Applied Mathematics, University of Waterloo, Waterloo, Ontario, N2L 3G1, Canada}
\affiliation{Perimeter Institute for Theoretical Physics, 31 Caroline St N, Waterloo, Ontario, N2L 2Y5, Canada}


\begin{abstract}
We use quantum energy teleportation in the light-matter interaction as an operational means to create quantum field states that violate energy conditions and have  negative local stress-energy densities. We show that the protocol is optimal in the sense that it scales in a way that saturates the quantum interest conjecture.
\end{abstract}

\maketitle

\section{Introduction}

In general relativity, energy conditions  are often used (and required) in proofs of important theorems about black holes and the existence  of spacetimes with singularities \cite{Hawking1972,hawking1973book}. Some of them capture the idea that energy and mass are both positive and place (in principle reasonable) restrictions on the allowed stress-energy tensors that source gravity in Einstein's equation. These energy conditions were found to be consistent with all forms of classical matter and had the additional `feature' of forbidding certain exotic solutions of Einstein's equation that posed serious mathematical and philosophical problems, such as naked singularities \cite{Hawking1972}, warp-drives \cite{Alcubierre94}, physically traversable wormholes \cite{Thorne88} and closed timelike curves \cite{visser96} (see  \cite{Curiel_2014} for a summary). In particular the weak energy condition can be understood as forbidding the existence of a `negative gravitational mass', i.e. a source of negative spacetime curvature, a concept which is supported by empirical evidence. These energy conditions, however, are known to be violated by quantum fields \cite{Casimir48,Ford93}, opening the door to the possible existence of more exotic spacetimes where gravity is sourced by quantum matter. 

The creation of non-classical states that violate classical energy conditions can, in principle, be performed in a laboratory setting (not without challenge) through the Casimir effect \cite{Casimir48,Casimir_exp_rev} and the generation of squeezed states \cite{Ford93,squeeze_review}.  The ability of quantum fields to violate classical energy conditions has lead to the formulation of the so called \textit{quantum energy conditions} \cite{Ford78,Pfenning98}, which have been derived from quantum inequalities and are inviolable by any state of matter described by relativistic quantum theory. These quantum energy conditions have been used to place restrictions on how much the classical energy conditions can be violated in certain energy density distributions such as those generated by the dynamical Casimir effect \cite{Fulling393,Wilson2011}.

In this paper we analyze a different operational protocol that allows us to engineer violations of the classical energy conditions using the light-matter interaction. That is, we explore the possibility that particle detectors coupled locally to a quantum field (one can think of atomic probes, for instance) can generate distributions of negative energy that saturate the quantum inequalities and allow for a reasonable degree of control of their spatial and temporal distribution. In particular we will use a protocol called  Quantum Energy Teleportation \cite{Hotta2008}.

Quantum energy teleportation (QET) was first proposed by Hotta \cite{QET_SPINS}, as a means to transport energy between two parties Alice and Bob, without the need for energy carriers to physically travel from Alice to Bob. It was first applied to quantum fields in \cite{Hotta2008}. There is an extensive amount of literature on QET  focusing on how underlying correlations in a quantum field state (such as, e.g., the vacuum) allow for energy transfer without energy carriers being exchanged. For example, it has been studied how the protocol efficacy can be improved by optimizing the field state to overcome the limitations placed by quantum energy conditions \cite{Hotta2010,Hotta2010b,Hotta2011xj,Hotta2014,Hotta2014,Verdon2016,Verdon_thesis,Hotta_Squeeze}. QET has also been used to demonstrate `stimulated Hawking radiation' \cite{Hotta2010c} where a black hole can be made to lose mass via a QET procedure, independent of the Hawking radiation already present. Related to this last work, our interest in QET is as a tool to `mold' particular stress-energy densities that violate  classical energy conditions.

The particular version of QET that we analyze in this paper is based on particle detectors coupled locally to scalar fields which have the ability to communicate in some form (classical or quantum) in regimes where these detectors are good models for the atom-light interaction \cite{Alhambra2013,Pozas2016}.

We will analyze this setup both in a toy 1+1 dimensional scenario and in the full 3+1 dimensional case, showing how much control it allows over the distribution of energy density that can be generated. We will show that, with this protocol, it is possible to construct negative energy densities that optimally saturate  quantum inequalities. The resources utilized by the protocol to generate custom-shaped negative energy distributions are arguably not demanding as a matter of principle, since the particle detectors can be thought of as inertial atomic clouds prepared in the ground state and interacting pointlikewise with the field in precisely the same way that hydrogenoid atoms interact with the electromagnetic field, as it is discussed in detail in \cite{Pozas2016}.

An outline of this paper follows: In section~\ref{sec2} we introduce the QET protocol used and outline the derivation for the final stress-energy tensor that then describes the negative energy distribution of the field in 1+1~D and 3+1~D. In section~\ref{sec3} we analyze the results and optimize the energy density of the field after the protocol is performed for both  1+1~D and 3+1~D. In section~\ref{sec4} we derive a scaling relation for the stress-energy density generated with this protocol and discuss its relation to the quantum interest conjecture \cite{Ford99}. Finally, in appendix~\ref{seca1} we give a detailed derivation of the final stress-energy tensor after the QET protocol is performed and also demonstrate how the optimal QET protocol we use is independent of the choice of quantum or classical communication. In particular we show the equivalence of a single non-local `Bob' and a cloud of multiple local `Bob's' regarding the stress-energy density produced with this protocol.

\section{Quantum Energy Teleportation in 1+1~D and 3+1~D}\label{sec2}

In the usual QET scenario introduced in \cite{Hotta2008}, Alice operates a particle detector that she couples locally to the field, gaining information about it. Then, this information is transmitted either through a classical channel \cite{Hotta2008} or a quantum one \cite{Verdon2016} to another observer Bob, who operates a different particle detector, and makes use of the information transmitted by Alice to break the local passivity of the field vacuum state, thus extracting energy from it \cite{Hotta2011xj}. 

In a large number of previous works analyzing QET protocols (mainly in 1+1~D) the particular setup analyzed involved placing Bob to the `right' of Alice, and adapting Alice's and Bob's particle detectors to couple exclusively to left-moving modes of the field (i.e. Alice's perturbation propagates away from Bob). These directional detectors were used to show in a clear way that energy does not travel from Alice to Bob, and yet Bob can extract energy from the field. 

In contrast, in this paper we are interested in the potential that atomic particle detectors may have to create and shape arbitrary negative energy distributions. If we want to make the setting resemble atomic-light interactions, there are some subtleties that need to be addressed. To begin with,  light-atom interactions do not violate parity. In particular it is not possible to couple a Hydrogen-like atom exclusively to modes propagating in a particular direction. Therefore the formalism that assumes such couplings, as analyzed in most previous literature (with the notable exception of \cite{Hotta2010b} where harmonic oscillators were coupled dipolarly to the electric field to establish a QET protocol), will not be sufficient for our purposes. In particular, we need to analyze QET protocols with detectors coupling to modes propagating in all directions in a more thorough way than in previous studies in order to give explicit forms for the exact energy density distribution that the protocol can generate.

The fact that we need to consider isotropic couplings poses further challenges when we want to use QET to create custom-shaped negative-energy distributions. Concretely, with an isotropic coupling, Alice's perturbation of the field will also propagate towards Bob. This in turn means that, when he interacts with the field (to extract energy from it and generate negative energy densities), he not only has to reduce the expectation value of the stress-energy density in his surroundings from zero to negative values, he also has to overcome whatever energy excess Alice created through her interaction that also propagates to Bob's surroundings. This is particularly relevant in the 3+1~D case as we will see below.

Our setup can be thought of as Alice having a single qubit (for example, a detector, or an atom), acting with it on the field and classically communicating her procedure to Bob's agents, operating an atomic cloud of many detectors (or atoms) surrounding Alice. 

For these purposes, it is important to note that the use of quantum communication or classical communication to send the information from Alice to Bob in a QET protocol does not make a difference in the yield of the protocol. In fact, it was shown in \cite{Verdon2016,Verdon_thesis} that both LOCC and LOQC based (qubit) QET protocols are equivalent in yield. 

More relevant to our case, in this paper we will present a proof that optimal LOQC QET where Bob operates a non-local qubit yields the same energy density distribution in the quantum field as an optimal LOCC QET protocol where Bob operates a `cloud' of local qubits  (this is done in detail in appendix~\ref{seca1}). This will allow us to use the LOQC formalism developed in \cite{Verdon2016} with Alice communicating through a quantum channel with a single (non-local) Bob to study the negative energy created by Alice communicating classically with a distribution of multiple Bobs. Note, that this is just for computational simplicity: the availability of a quantum communication channel from Alice to Bob, or the non-locality of Bob, are not necessary requirements for our analysis  nor do they affect the validity of our results, as we will show in full detail.

\subsection{QET in 1+1}

Let us begin with the simple case of 1+1~D. We will follow the standard LOQC QET protocol outlined in \cite{Verdon2016} using qubits rather than general qudits as our detectors. For the reasons exposed above, we also modify the published protocol to consider detector-field couplings that do not break left-right symmetry. 

Let us consider two observers Alice and Bob operating particle detectors that can be switched on and off in a controlled manner. Alice's detector is a qubit (two-level quantum system) that couples to the field via a $\delta$-switching at an arbitrary time $t_{0}$ through the following Unruh-DeWitt derivative coupling Hamiltonian
\begin{align}
\hat{H}_{\textsc{i}}(t)=&\delta(t-t_{0})\hat{\sigma}_{x}\int\limits_{\mathbb{R}} \dd x\, \lambda(x)\hat{\pi}(x).\label{AliceH}
\end{align}

 Following this interaction Alice's detector is sent to Bob via some quantum channel in the same fashion as in \cite{Verdon2016}. For example, Alice could have teleported her detector's state  to Bob's  detector \cite{Wootters93}.

After this, Bob  allows the detector to interact with the field at a later time $T$ via the Unruh-DeWitt coupling (UDW)
\begin{align}
\hat{H}_{\textsc{i}}(t)=&\delta(t-T-t_{0})\hat{\sigma}_{z}\int\limits_{\mathbb{R}}\dd x \,\mu(x)\hat{\phi}(x).\label{BobH}
\end{align}
Notice that the couplings \eqref{AliceH} and \eqref{BobH} capture most of the fundamental aspects of the light-matter interaction in absence of exchange of angular momentum \cite{Alhambra2013,Pozas2016}. The smearing functions $\lambda$ and $\mu$ describe the detectors' interaction region. $\lambda,\mu$ and time $T$ must be chosen such that Bob only receives the information (and uses it) when he enters the lightcone of Alice. The smearing functions are further assumed to have compact supports and have continuous second derivatives on this support. The use of the $\delta$ function for temporal switching allows us to treat the problem non-perturbatively in the coherent state basis.

Note that Alice's and Bob's observables coupled to the field in \eqref{AliceH} and \eqref{BobH} correspond to non-commuting generators of SU(2) ($\hat \sigma_x,\hat\sigma_z$). This is crucial for the QET protocol to work. The choices of derivative coupling for Alice and regular UDW coupling for Bob are just taken for convenience as in \cite{Verdon2016}. 

We will analyze the energy density in the scalar field at the different stages of the protocol. The scalar field stress-energy tensor is given by N\"{o}ther's theorem as
\begin{align}
\hat{T}_{\mu\nu}=\partial_{\mu}\hat{\phi}\partial_{\nu}\hat{\phi}-\eta_{\mu\nu}\left(\partial_{\rho}\partial^{\rho}\hat{\phi}\right).
\end{align}
We will be looking at the $\hat T_{00}$ component of the normal ordered stress-energy tensor ($:\hat{T}_{00}(x):$) corresponding to `energy density'. When $:\hat{T}_{00}(x):$ is integrated over all space we recover the scalar field Hamiltonian of decoupled momentum harmonic oscillators. This tells us that if we have local negative energy densities (i.e. negative expectation values of $:\hat{T}_{00}(x):$) then they must be offset by positive energy densities elsewhere in order to satisfy non-negativity of the renormalized field Hamiltonian. 

Working in the Schr\"{o}dinger picture (and taking w.l.o.g. $t_{0}=0$), the state of the system is initially $\ket{\psi}=\ket{0}\otimes\ket{A_{0}}$ where $\ket{0}$ is the Minkowski vacuum of the field and $\ket{A_{0}}$ is the initial state of the detector. Solving Schr\"{o}dinger's equation with the full interaction Hamiltonian leads to the following state after Alice's interaction:
\begin{align}
\nonumber \ket{\psi(t)}&= e^{-\ii \hat{\sigma}_{x}\int\limits_{\mathbb{R}} \dd x\lambda(x) \hat{\pi}(x)}\ket{0}\ket{A_{0}}\\
\nonumber =&\exp\left(-\hat{\sigma}_{x}\int \dd k \,\sqrt{\frac{\abs{k}}{4\pi}}\left[\hat{a}_{k}^{\phantom{\dagger}}\tilde{\lambda}^{*}(k)-\hat{a}_{k}^{\dagger}\tilde{\lambda}(k)\right]\right)\ket{0}\ket{A_{0}}\\
\nonumber =&\exp\left(\hat{\sigma}_{x}\int \dd k \left[\alpha_{k}^{\phantom{*}}\hat{a}_{k}^{\dagger}-\alpha_{k}^{*}\hat{a}_{k}^{\phantom{\dagger}}\right]\right)\ket{0}\ket{A_{0}}\\
 =&\ket{\bm{\alpha}(t)}\ket{\px}\braket{\px|A_{0}}+\ket{-\bm{\alpha}(t)}\ket{\mx}\braket{\mx|A_{0}}.\label{eq6}
\end{align}
for $0<t<T$. Note that, because of the delta coupling, the expectation of the stress-energy tensor does not depend on the energy gap of the detector, and as such, the contribution of free Hamiltonian $\Omega \hat{\sigma}^{+}\hat{\sigma}^{-}$ plays no role in this protocol. In \eqref{eq6}, $\exp\left(\pm \dd k \left[\alpha_{k}^{\phantom{*}}\hat{a}_{k}^{\dagger}-\alpha_{k}^{*}\hat{a}_{k}^{\phantom{\dagger}}\right]\right)$ is the (momentum) coherent state displacement operator with displacement vector $\alpha_{k}$ such that, when acting on the vacuum, the resulting state obeys $\hat{a}_{k}\ket{\alpha_{k'}}=\delta_{kk'}\alpha_{k}\ket{\alpha_{k}}$, $\braket{0|\alpha_{k}}=e^{-\dd k\frac{\abs{\alpha_{k}}^{2}}{2}}$. Here $\ket{\bm{\alpha}(t)}$ corresponds to a tensor product of momentum coherent states
\begin{align}
\ket{\bm{\alpha}(t)}=&\bigotimes_{k\in \mathbb{R}}\ket{\alpha_{k}(t)},\\
\alpha_{k}(t)=&e^{-\ii\left|k\right| t}\sqrt{\frac{\left|k\right|}{4\pi}}\int \dd x\, \lambda(x) e^{-\ii kx}.
\end{align} 
This state has an energy density given by
\begin{align}
\left<:\hat{T}_{00}(x,t):\right>=&\frac{1}{4}\left(\lambda'(x-t)\right)^{2}+\frac{1}{4}\left(\lambda'(x+t)\right)^{2}.\label{eq8}
\end{align}
Alice's interaction has resulted in an injection of energy into the field that propagates away from Alice at the speed of light. Additionally, this interaction has not introduced negative energy densities of any sort. Alice's interaction Hamiltonian was dependent on $\hat{\pi}$, consequently, the input energy depends on the first derivative of Alice's smearing function. Note that when $t=0$ the stress-energy tensor is non-zero only on the support of the smearing function, which is to be expected due to the local nature of the interaction.

Following Bob's interaction ($t>T$) the system is in the state
\begin{align}
\begin{split}
&\ket{\psi(t)}=\\
&\hat{D}(\bm{\beta}(t))\left(\frac{\braket{\px|A_{0}}}{\sqrt{2}}\ket{\bm{\alpha}(t)}\ket{\pz}+\frac{\braket{\mx|A_{0}}}{\sqrt{2}}\ket{-\bm{\alpha}(t)}\ket{\pz}\right)\\
+&\hat{D}(-\bm{\beta}(t))\left(\frac{\braket{\px|A_{0}}}{\sqrt{2}}\ket{\bm{\alpha}(t)}\ket{\mz}-\frac{\braket{\mx|A_{0}}}{\sqrt{2}}\ket{-\bm{\alpha}(t)}\ket{\mz}\right),
\end{split}\label{eq10}
\end{align}
where
\begin{align}
&\beta_{k}(t)=-\frac{\ii e^{-\ii\left|k\right|(t-T)}}{\sqrt{4\pi\left|k\right|}}\int \dd x\, \mu(x)e^{-\ii kx},\\
&\hat{D}(\bm{\beta}(t))=\exp\left(\int\dd k\,\left[\beta_{k}^{\vphantom{*}}(t)\hat{a}_{k}^{\dagger}-\beta_{k}^{*}(t)\hat{a}_{k}^{\vphantom{\dagger}}\right]\right),
\end{align}
where $\hat{D}(\bm{\beta})$ is a coherent state displacement operator.
When taking expectation values of the stress-energy tensor with the post-Bob state we can see that each individual coherent state e.g. $\hat{D}(\bm{\beta}(t))\ket{\bm{\alpha}(t)}$ will have a positive contribution to the stress-energy tensor, similar to the case of \eqref{eq8}. It is the cross terms e.g. $\braket{\bm{\alpha}(t)|\hat{D}^{\dagger}(\bm{\beta}(t)):\hat{T}_{00}(x,t):\hat{D}(\bm{\beta}(t))|-\bm{\alpha}(t)}$ that will become responsible for the generation of negative energy densities. These non-trivial cross-terms will be collectively referred to as the QET term herein. The very nature of the QET term as a cross term highlights the importance of coherence in this protocol.

The post QET state's energy density is more challenging to write down,
\begin{widetext}
\begin{align}
\begin{split}
&\left<:\hat{T}_{00}(x,t):\right>=\frac{\left(\lambda'(x-t)\right)^{2}}{4}+\frac{\left(\lambda'(x+t)\right)^{2}}{4}+\frac{\left(\mu(x-(t-T))\right)^{2}}{4}+\frac{\left(\mu(x+(t-T))\right)^{2}}{4}\\
+&\underbrace{\frac{e^{-2\|\alpha\|}\braket{A_{0}|\hat{\sigma}_{y}|A_{0}}}{2\pi}\mu(x-(t-T))\int \dd y\, \lambda'(y)\frac{\text{P.P}}{y-x+t}}_{\text{Right moving QET term}}
+\underbrace{\frac{e^{-2\|\alpha\|}\braket{A_{0}|\hat{\sigma}_{y}|A_{0}}}{2\pi}\mu(x+(t-T))\int \dd y\, \lambda'(y)\frac{\text{P.P}}{y-x-t}}_{\text{Left moving QET term}}.
\end{split}\label{eq11}
\end{align}
\end{widetext}
Where $\|\alpha\|=\int \dd k\left|\alpha_{k}\right|^{2}$, and $\text{P.P}$ denotes Cauchy's principal value. These integrals are of the same nature as those previously reported in the literature of QET \cite{Verdon2016}. Here we have managed to isolate the QET term from the positive energy density contributions. In \eqref{eq11} the terms $\frac14\left(\lambda'(x\pm t)\right)^{2}$ denote the contributions to the field energy by Alice's detector-field interaction and similarly $\frac14\left(\mu(x\pm(t-T))\right)^{2}$ is Bob's detector-field interaction contribution to the energy density. The QET term is the only one that involves both smearing functions and is dependent on the ground state entanglement, the coherence of the system during the course of the protocol and the fidelity of the quantum channel \cite{Hotta_Squeeze,Hotta2010_intro}. Given that it is linear in $\lambda'$ we can find smearing functions that allow this QET term to be negative enough to overcome the effects of both Alice's and Bob's energy contribution to the field. 

It is important to note that whilst the QET protocol can be used to extract energy from the field, one could equally use a QET-like protocol to efficiently inject energy into the field. Namely, if we choose $\ket{A_{0}}$ appropriately we can increase the energy density to beyond the sum of Alice's and Bob's independent contributions. The freedom of choosing the initial detector configuration is one of the few ways to control the protocol.

Recall that we are studying our ability to shape field energy distributions using QET by manipulating Alice's and Bob's smearings. Let us look at \eqref{eq11}, and in particular at the integral in the QET terms. This principal value integral can be written as
\begin{align}
\int \dd y\, \lambda'(y)\frac{\text{P.P}}{y-x+t}=&\lim_{\varepsilon\rightarrow 0}\Bigg[\int\limits^{x-t-\varepsilon}_{-\infty}+\int\limits_{x-t+\varepsilon}^{\infty}\Bigg]\dd y \frac{\lambda'(y)}{y-x+t}{\color{blue}.}{\color{red},}
\end{align}
where we have introduced the abbreviated notation 
\begin{equation}
    \Bigg[\int\limits^{b}_{a}+\int\limits_{c}^{d}\Bigg]\dd x\, f(x) \coloneqq  \int\limits^{b}_{a}\dd x\, f(x)+\int\limits_{c}^{d}\dd x\, f(x).
\end{equation}

Evaluating this principal value integral numerically poses a relatively challenging problem around the singular value of the denominator. In order to do so in an efficient way, we will further subdivide the integration domain as follows: $(-\infty,x-t-a],(x-t-a,x-t-\varepsilon],[x-t+\varepsilon,x-t+a),[x-t+a,\infty)$ where $a$ is some positive real number
\begin{align}
\begin{split}
\int \dd y \,\lambda'(y)\frac{\text{P.P}}{y-x+t}=&\Bigg[\int\limits^{x-t-a}_{-\infty}+\int\limits_{x-t+a}^{\infty}\Bigg]\dd y \frac{\lambda'(y)}{y-x+t}\\
+&\lim_{\varepsilon\rightarrow 0}\Bigg[\,\int\limits_{x-t-a}^{x-t-\varepsilon}+\int\limits^{x-t+a}_{x-t+\varepsilon}\,\Bigg]\dd y \frac{\lambda'(y)}{y-x+t}.
\end{split}
\end{align}
In the integration domains $(x-t-a,x-t-\varepsilon],[x-t+\varepsilon,x-t+a)$ we will further assume that we can expand $\lambda'(y)$ using Taylor's theorem around $y=x-t$ in order to deal with the pole:
\begin{align}
\begin{split}
&\int \dd y\, \lambda'(y)\frac{\text{P.P}}{y-x+t}=\Bigg[\int\limits^{x-t-a}_{-\infty}+\int\limits_{x-t+a}^{\infty}\Bigg]\dd y \frac{\lambda'(y)}{y-x+t}\\
+&\lim_{\varepsilon\rightarrow 0}\Bigg[\,\int\limits_{x-t-a}^{x-t-\varepsilon}+\int\limits^{x-t+a}_{x-t+\varepsilon}\,\Bigg]\dd y \left[\frac{\sum_{n=0}^{2}\lambda^{(n+1)}(x-t)\frac{(y-x+t)^{n}}{n!}}{y-x+t}\right.\\
&+\left.\frac{\lambda^{(4)}(\xi)\frac{(y-x+t)^{3}}{3!}}{y-x+t}\right],
\end{split}
\end{align}
where $\xi$, between $x-t$ and $y$, bounds the error.

Following the Taylor expansion around $y=x-t$ we now commute the sum and integral and analytically evaluate the principal value integral resulting in
\begin{align}
\begin{split}
&\int \dd y\, \lambda'(y)\frac{\text{P.P}}{y-x+t}=\Bigg[\int\limits^{x-t-a}_{-\infty}+\int\limits_{x-t+a}^{\infty}\Bigg]\dd y \frac{\lambda'(y)}{y-x+t}\\
+& 2a\lambda''(x-t)+O\left(\lambda^{(4)}(\xi)\right)\frac{a^{3}}{9}.
\end{split}
\end{align}
Note that this can be done due to the fact that the principal value integrals are not divergent in the first place (yet difficult to evaluate numerically without using these tools). Since $\abs{\lambda^{(4)}(\xi)}$ is bounded on the domain of interest we can choose $a$ small enough (the particular value will depend on the particular choice of smearing) such that we only need to keep the first term of the Taylor polynomial
\begin{align}
&\int \dd y\, \lambda'(y)\frac{\text{P.P}}{y-x+t}=\lim_{\varepsilon\rightarrow 0}\Bigg[\int\limits^{x-t-a}_{-\infty}+\int\limits_{x-t+a}^{\infty}\Bigg]\dd y \frac{\lambda'(y)}{y-x+t}\nonumber\\
&\qquad\quad+2a\lambda''(x-t)+\mathcal{O}\Big(a^3 \lambda^{(4)}(\xi)\Big).
\label{eq17}
\end{align}
We will use this approximation only for numerical evaluation later on in the paper (after checking that the higher order terms  provide a negligible contribution with respect to the $a\lambda''$ term). 

All this considered, we see that to maximize the QET term in \eqref{eq11} with respect to Alice's local term, we are looking for a smearing function that has a large second derivative and small first derivative (so that Alice's contribution to the field energy density is not too high). Whilst it is not possible to have a smearing function that obeys this criteria over the whole support, we know that a large second derivative and small first derivative occurs around functions' maxima. Examples of functions with regions where this is so that are easy to visualize are Gaussians and Lorentzians. The fact that the second derivative of smooth functions is larger than the first derivative will happen only around maxima (and not in the whole domain) tells us that any negative energy density will be accompanied by positive energy and the larger we make the region of negative energy the shallower this region will be. These mathematical restrictions limit the amount of isolated negative energy packets, and are in fact connected with the inability to violate quantum energy conditions (e.g. Flanagan's theorem \cite{Flanagan97}), as we will discuss in more detail in later sections.

\subsection{QET in 3+1}

The QET protocol presented above can be generalized to 3+1~D in a relatively straightforward way, allowing us to study the creation of negative energy densities in 3+1~D. The full interaction Hamiltonian describing the protocol resembles the 1+1~D case:
\begin{align}
\nonumber \hat{H}_{\textsc{i}}(t) &= \delta(t)\hat{\sigma}_{x}\int\limits_{\mathbb{R}^{3}} \dd^{3}\bm{x}\, \lambda(\bm{x})\hat{\pi}(\bm{x})\\&\qquad+\delta(t-T) \hat{\sigma}_{z}\int\limits_{\mathbb{R}^{3}}\dd^{3}\bm{x}\, \mu(\bm{x})\hat{\phi}(\bm{x}).
\end{align}

Again the smearing functions and $T$ must be chosen so that they do not violate causality. Note that, as explained above, we will be using the LOQC variant of the protocol only as a mathematical tool to calculate the distribution of negative energy obtained via a delocalized LOCC variant of QET. Recall that, as shown in detail in appendix~\ref{seca1}, an LOQC protocol operating from Alice to a non-local  single Bob (with spatial distribution $\mu$) is exactly equivalent to an LOCC protocol from a single Alice to a distribution ($\mu$) of many local Bobs.

Alice's and Bob's interactions proceed in the same manner as the 1+1~D case resulting in a final state analogous to \eqref{eq10}:
\begin{align}
\begin{split}
&\ket{\psi(t)}=\\
&\hat{D}(\bm{\beta}(t))\left(\frac{\braket{\px|A_{0}}}{\sqrt{2}}\ket{\bm{\alpha}(t)}\ket{\pz}+\frac{\braket{\mx|A_{0}}}{\sqrt{2}}\ket{-\bm{\alpha}(t)}\ket{\pz}\right)\\
+&\hat{D}(-\bm{\beta}(t))\left(\frac{\braket{\px|A_{0}}}{\sqrt{2}}\ket{\bm{\alpha}(t)}\ket{\mz}-\frac{\braket{\mx|A_{0}}}{\sqrt{2}}\ket{-\bm{\alpha}(t)}\ket{\mz}\right),
\end{split}\\
&\alpha_{\bm{k}}(t)=e^{-\ii\left|\bm{k}\right| t}\sqrt{\frac{\left|\bm{k}\right|}{4\pi}}\frac{1}{2\pi}\int \dd^{3} \bm{x}\, \lambda(\bm{x}) e^{-\ii \bm{k}\cdot\bm{x}},\\
&\beta_{\bm{k}}(t)=-\frac{\ii e^{-\ii\left|\bm{k}\right|(t-T)}}{2\pi\sqrt{4\pi\left|\bm{k}\right|}}\int \dd^{3} \bm{x}\, \mu(\bm{x})e^{-\ii \bm{k}\cdot\bm{x}},\\
&\hat{D}(\bm{\beta}(t))=\exp\left(\int\dd^{3} \bm{k}\,\left[\beta_{\bm{k}}^{\vphantom{*}}(t)\hat{a}_{\bm{k}}^{\dagger}-\beta_{\bm{k}}^{*}(t)\hat{a}_{\bm{k}}^{\vphantom{\dagger}}\right]\right).
\end{align}

The final energy density can be decomposed in a similar way to the 1+1~D case, although it is inherently more complicated,
\begin{widetext}
\begin{align}
\begin{split}
&\braket{\psi\left(T+\Delta T\right)|:\hat{T}_{\mu\nu}:|\psi\left(T+\Delta T\right)}=
\frac{1}{4^{4}\pi^{6}}
\bigg\{\underbrace{\bigg(I_{\mu}^{1}I_{\nu}^{1}-\frac{\eta_{\mu\nu}}{2}I_{\lambda}^{1}I^{1\lambda }\bigg)}_{\text{Bob's field contribution}}-\underbrace{\bigg(I_{\mu}^{2}I_{\nu}^{2}-\frac{\eta_{\mu\nu}}{2}I_{\lambda}^{2}I^{2\lambda }\bigg)}_{\text{Alice's field contribution}}\\
-&\underbrace{e^{-2\|\alpha\|}\braket{A_{0}|\hat{\sigma}_{y}|A_{0}} \bigg(I_{\mu}^{1}I_{\nu}^{3}-\frac{\eta_{\mu\nu}}{2}I_{\lambda}^{1}I^{3\lambda}\bigg)
-e^{-2\|\alpha\|}\braket{A_{0}|\hat{\sigma}_{y}|A_{0}} \bigg(I_{\mu}^{3}I_{\nu}^{1}-\frac{\eta_{\mu\nu}}{2}I_{\lambda}^{3}I^{1\lambda}\bigg)}_{\text{QET terms}}\bigg\}.
\end{split}\label{eq50}
\end{align}

W.l.o.g. Alice performs her operation at $t=0$, Bob performs his operation at $t=T$ and we are observing the energy density at $t=T+\Delta T$. The energy density from Alice alone can be read from \eqref{eq50} by merely setting $\mu= 0$ everywhere. The terms $I^{i}_{\mu}$ are defined as
\begin{align}
I^{1}_{\mu}=&\int \dd ^{3}\bm{r} d^{3}\bm{k}\,\hat{k}_{\mu}
\bigg(
e^{\left|\bm{k}\right|\left(-2\varepsilon+\ii\Delta T\right) +\ii\bm{k}\cdot\left(\bm{r}-\bm{x}\right)}
+e^{\left|\bm{k}\right|\left(-2\varepsilon-\ii\Delta T\right) -\ii\bm{k}\cdot\left(\bm{r}-\bm{x}\right)}\bigg)\mu\left(\bm{r}\right),\\
I^{2}_{\mu}=&\int \dd ^{3}\bm{r} d^{3}\bm{k}\,\hat{k}_{\mu}
\bigg(
e^{\left|\bm{k}\right|\left(-2\varepsilon-\ii(\Delta T+T)\right) -\ii\bm{k}\cdot\left(\bm{r}-\bm{x}\right)}-e^{\left|\bm{k}\right|\left(-2\varepsilon+\ii(\Delta T+T)\right) +\ii\bm{k}\cdot\left(\bm{r}-\bm{x}\right)}\bigg)\left|\bm{k}\right|\lambda\left(\bm{r}\right),\\
I^{3}_{\mu}=&\int \dd ^{3}\bm{r} d^{3}\bm{k}\,\hat{k}_{\mu}
\bigg(
e^{\left|\bm{k}\right|\left(-2\varepsilon-\ii(\Delta T+T)\right) -\ii\bm{k}\cdot\left(\bm{r}-\bm{x}\right)}+e^{\left|\bm{k}\right|\left(-2\varepsilon+\ii(\Delta T+T)\right) +\ii\bm{k}\cdot\left(\bm{r}-\bm{x}\right)}\bigg)\left|\bm{k}\right|\lambda\left(\bm{r}\right),
\end{align}
\end{widetext}

where $\|\alpha\|=\int \dd^{3}\bm{k}\abs{\alpha_{\bm{k}}}^{2}$.  

 To evaluate these expressions we restrict ourselves to linear combinations of spherically symmetric functions centered respectively in $\bm{x}_{\textsc{a}}$ and $\bm{x}_{\textsc{b}}$ for $\lambda$ and $\mu$. Even under this simplifying assumption of spherical symmetry no pleasant analytic solution presents itself and so these expressions will need to be evaluated numerically. Very coarsely, Alice's contribution to the field will be a function of $\partial_{r}(r\lambda(r))$ and the QET term will be proportional to $\partial_{r}^{2}(r\lambda(r))$.

\section{Creating negative energy densities}\label{sec3}

\subsection{1+1~D}

As was shown above, there are 3 terms of importance contributing to the total energy density; Alice's contribution and Bob's contribution both of which are positive and the QET terms that are the root of the energy teleportation protocol. In particular we have Alice's contribution proportional to $(\lambda')^{2}$, Bob's contribution proportional to $\mu^{2}$ and the QET terms, dominated by a term proportional to $\mu \lambda''$. These proportionality relations allow for some intuition on the numerical results. Namely, it suggests that we choose $\lambda$ and $\mu$ such that the support of $\mu$ coincides with the region where $\lambda'$ is small and $\lambda''$ is largest. 

\subsubsection{Method}

Three different parametrized smooth smearing functions $f,g$ and $h$ were used, in particular we use the smooth compactly supported function used in previous literature \cite{A_func}  
\begin{align}
\nonumber &f(z,\sigma,\delta)\\
=&\begin{cases}
S\left(\frac{\frac{\sigma}{2}+\pi\delta+z}{\delta}\right) & \text{if } -\pi\delta<z+\frac{\sigma}{2}<0\\
1 &\text{if } -\frac{\sigma}{2}\leq z\leq \frac{\sigma}{2}\\
S\left(\frac{\frac{\sigma}{2}+\pi\delta-z}{\delta}\right) & \text{if } 0<z-\frac{\sigma}{2}<\pi \delta\\
0 & \text{otherwise},
\end{cases}
\label{eq26}
\end{align}
where $S(x)=\frac{1}{2}(1-\text{tanh }\text{cot}(x))$, and the two analytic functions \begin{align}
&g(z,\delta)=\frac{1}{ \sqrt{2\pi}}e^{-\frac{z^{2}}{2\delta^{2}}},\\
&h(z,\delta)=\frac{1}{\pi}\frac{1}{1+\left(\frac{z}{\delta}\right)^{2}},\label{eq27}
\end{align}
where $z=x-x_{0}$. That is we used a smooth class infinity bump function ($f$), a Gaussian ($g$) and a Lorentzian ($h$) to model the interaction domain of the detectors. We consider the following three cases:
\begin{enumerate}
    \item $\lambda(x)=\lambda_{0} f(x,\sigma_{\textsc{a}},\delta_{\textsc{a}})$, \quad $\mu(x)=\mu_{0} f(x-x_{\textsc{b}},\sigma_{\textsc{b}},\delta_{\textsc{b}})$
    \item $\lambda(x)=\lambda_{0} g(x,\delta_{\textsc{a}})$, \quad\quad  $\,\,\,\mu(x)=\mu_{0} g(x-x_{\textsc{b}},\delta_{\textsc{b}})$    \item $\lambda(x)=\lambda_{0} h(x,\delta_{\textsc{a}})$, \quad\quad  $\,\,\,\mu(x)=\mu_{0} h(x-x_{\textsc{b}},\delta_{\textsc{b}})$.
\end{enumerate}
In these three cases, we perform an optimization of the parameters $x_{\textsc{b}}$, $\sigma_{\textsc{a},\textsc{b}}$ and $\delta_{\textsc{a},\textsc{b}}$ and search for the maximum negative energy creation.

The approach we took to create a negative energy density region was to systematically assign some $\delta_{\textsc{a}}$ to Alice (along with $\sigma_{A}=0$ for the case of the smooth bump) and chose some region that we wished to be the negative energy well. Because of the intuition discussed above, we chose that region to be around Alice's perturbation wave (we will create net negative energy in the region of the pulse of positive energy generated by Alice's action that can be seen in Fig.~\ref{fig1}), and with a width being some arbitrary fraction of $\delta_{\textsc{a}}$. This region's energy was then made its most negative by optimizing over Bob's position $x_{\textsc{b}}$, width $\delta_{\textsc{b}},\sigma_{\textsc{b}}$ as well as the heights of the smearings $\lambda_{0},\mu_{0}$, thus yielding a set of optimal parameters. We will show what the results are in this section and we will later show how we can rescale those negative energy distributions  at will  in section~\ref{sec4}.

\subsubsection{Results}
\begin{figure}[!h]
\includegraphics[scale=1]{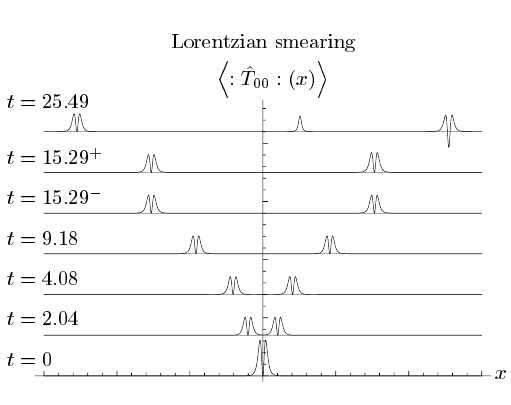}
\caption{Progression of the 1+1~D energy density wave is shown with each time slice offset in the y axis. As time progresses the wave can be seen propagating in both directions until $t=T=15.29$ when Bob's interaction introduces the negative energy density, obscured by a left moving positive energy contribution. Later propagation shows Bob's contribution splitting into left and right propagators leaving the desired significant negative energy density. Here a Lorentzian smearing was used.}\label{fig1}
\end{figure}

Following the determination of the optimal parameters for the negative energy well, the time evolution of the field could be simulated and compared to the intuition that can be derived from inspecting \eqref{eq11} and explained above. Fig.~\ref{fig1} shows the time evolution of the scalar field's energy density at several times, introducing an (uneven) offset in the y axis to represent time for illustration purposes. This plot displays the expected features of propagation in 1+1~D: there is no dispersion of the wavepacket and that there is both a left and right propagating wavepacket. For time $t=T$ two plots are shown, one before Bob's operation and one after to show the effects of QET on the energy density. On the support of Bob's smearing the energy density has changed and introduced a negative energy density region. By evolving the field further forward the negative energy region increases, since the left moving contribution of Bob's interaction moves away, as seen in Fig.~\ref{fig1}.

\begin{figure*}[t]
\includegraphics[scale=1]{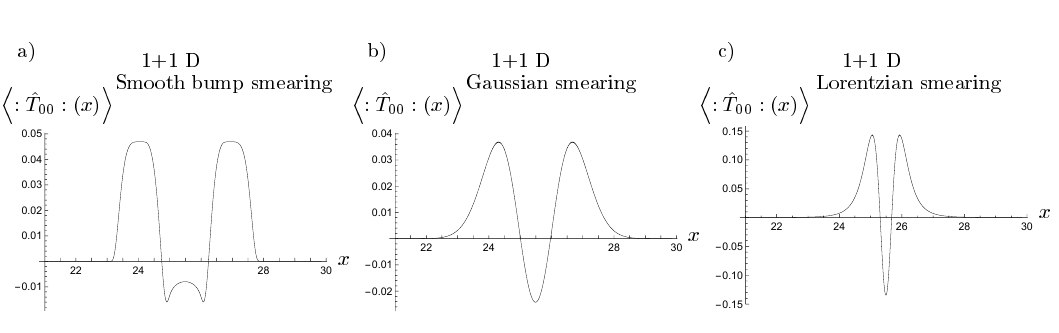}
\caption{Energy density at $x=T+\Delta T$, where Bob's interaction took place at $t=T$. The additional time $\Delta T$ is so the Bob's left moving energy packet moves away, as seen in Fig.~\ref{fig1}, increasing the amount of negative energy present. These 3 plots (from the left $f,g$ and $h$ smearings) support notion of large second derivatives relative to the first derivative of Alice's smearing $\lambda$.}\label{fig2}
\end{figure*}

From \eqref{eq17} we knew that the region of greatest negative energy density would correspond to that of the largest second derivative of Alice's smearing $\lambda$. This is reflected in Fig.~\ref{fig1} for the Lorentzian smearing, and is also displayed in Fig.~\ref{fig2} (some time after Bob interacted with the field). Of the three smearing functions, the Lorentzian has the sharpest peak, i.e. the largest second derivative, with respect to its length scale and this is seen in Fig.~\ref{fig2}c with the Lorentzian producing a very deep negative energy well with respect to  its adjacent positive energy peaks. Fig.~\ref{fig2} also reveals another aspect of 1+1~D QET that is irrespective of the choice of smearing functions i.e., once optimized, the negative energy well will sit between two positive energy peaks. The protocol could also be adjusted so that a negative energy well appears behind the wavepacket (although obtaining a much smaller amount of negative energy). Note that if the field is massless then a negative energy well in front of the wavepacket is not possible due to constraints imposed by causality, i.e. Bob would have to be outside of Alice's lightcone.

\subsection{3+1~D}
In 3+1~D the energy density \eqref{eq50} has three relevant terms, same as in the 1+1~D case. However these terms $I_{\mu}^{i}$ are significantly more complicated to solve than in 1+1~D. To simplify the analysis of the 3+1~D case we will make extra simplifying assumptions on the smearing functions $\lambda$ and $\mu$. In particular, we will assume that $\lambda$ and $\mu$ are spherically symmetric around some arbitrary point in space (not necessarily the same point). 

\subsubsection{Method}

The smearings of Alice and Bob that we used were spherically symmetric around $\bm{x}_{\textsc{a}}$ and $\bm{x}_{\textsc{b}}$ respectively, i.e. $\lambda(\bm{x})=\lambda\left(\abs{\bm{x}-\bm{x}_{\textsc{a}}}\right)$ and $\mu(\bm{x})=\mu\left(\abs{\bm{x}-\bm{x}_{\textsc{b}}}\right)$. As particular cases for the numerical analysis we tested the same three families of parametrized smearing functions used in the 1+1~D case, that is, those given by eq.\eqref{eq26}-\eqref{eq27} where $z_{\textsc{a}}=\abs{\bm{x}-\bm{x}_{\textsc{a}}}-r^{\textsc{A}}_{0}$, $z_{\textsc{b}}=\abs{\bm{x}-\bm{x}_{\textsc{b}}}-r_{0}^{\textsc{B}}$. Namely, 
\begin{enumerate}
    \item $\lambda(\bm x)=\lambda_{0} f(z_{\textsc{a}},\sigma_{\textsc{a}},\delta_{\textsc{a}})$, \quad $\mu(\bm x)=\mu_{0} f(z_{\textsc{b}},\sigma_{\textsc{b}},\delta_{\textsc{b}})$
    \item $\lambda(\bm x)=\lambda_{0} g(z_{\textsc{a}},\delta_{\textsc{a}})$, \quad\quad  $\,\,\,\mu(\bm x)=\mu_{0} g(z_{\textsc{b}},\delta_{\textsc{b}})$    \item $\lambda(\bm x)=\lambda_{0} h(z_{\textsc{a}},\delta_{\textsc{a}})$, \quad\quad  $\,\,\,\mu(\bm x)=\mu_{0} h(z_{\textsc{b}},\delta_{\textsc{b}})$.
\end{enumerate}

\begin{figure}[!h]
\includegraphics[scale=1]{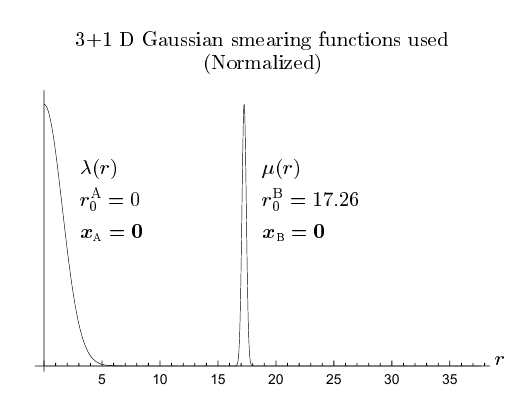}
\caption{Radial slice of the (normalized) smearing functions used for 3+1~D Gaussian smearing QET. With $r_{0}^{\textsc{B}}\neq 0$ and $\bm{x}_{\textsc{a}}=\bm{x}_{\textsc{b}}$, $\mu(r)$ resembles a shell surrounding Alice.}\label{fig2_5}
\end{figure}

When $r^{\textsc{A},\textsc{B}}_{0}=0$ these smearing functions correspond to spherically symmetric functions whose strength decays with distance. When $r^{\textsc{A},\textsc{B}}_{0}\neq 0$ the smearing resembles a spherically symmetric shell. For our calculations we are going to have Alice and Bob centered around the same point with Alice $r_{0}^{\textsc{A}}=0$ and Bob $r_{0}^{\textsc{B}}>0$ (See Fig.~\ref{fig2_5} for illustration). This way, Alice's contribution would lead to an outgoing spherical wave that would eventually reach Bob's `shell'-like smearing and complete the QET procedure.  Recall that  Bob's non-locality as a shell is shown to be an (exact) way to model a cloud of many local Bob's in appendix~\ref{seca1} using LOCC=LOQC in qubit QET.

\subsubsection{Results}
\begin{figure*}[t]
\includegraphics[scale=1]{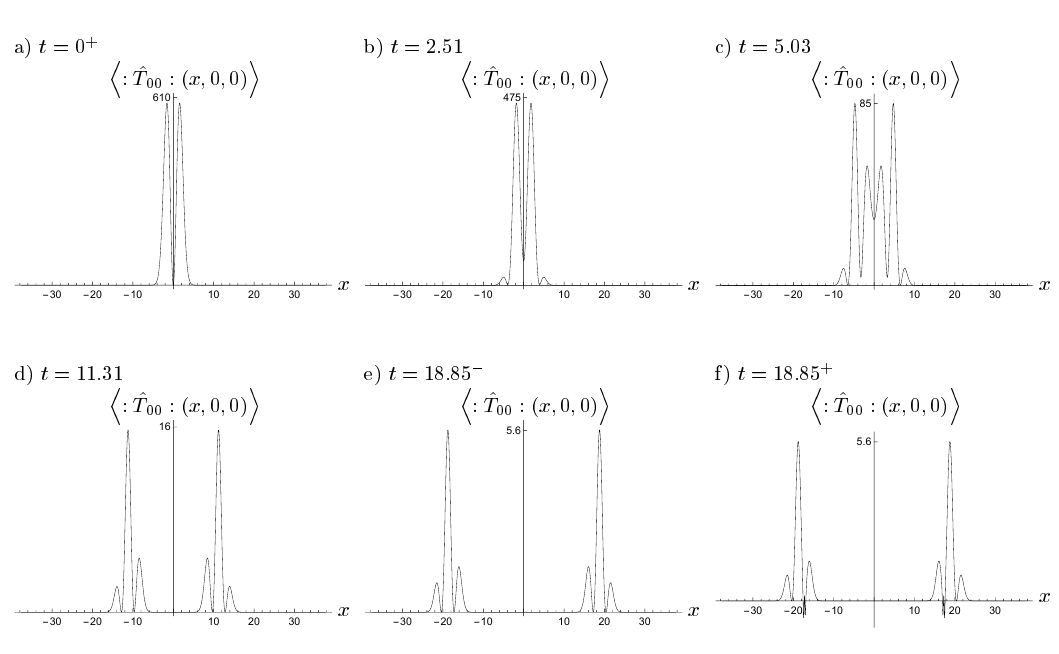}
\caption{$x$-axis slice of the progression of the 3+1~D energy density wave at various times with Bob's interaction taking place at $t=T=18.85$. The waves are `ingoing' and `outgoing' with the spherical nature of these waves apparent by their decay with increasing $\abs{x}$. Immediately following Bob's interaction there is a small amount of negative energy that is being partially obscured by Bob's ingoing positive energy contribution. Here a Gaussian smearing was used. We also refer the reader to Fig.~\ref{fig4}b, which shows the energy distribution following the separation of in and outgoing waves. }\label{fig3}
\end{figure*}

\begin{figure*}[t]
\includegraphics[scale=1]{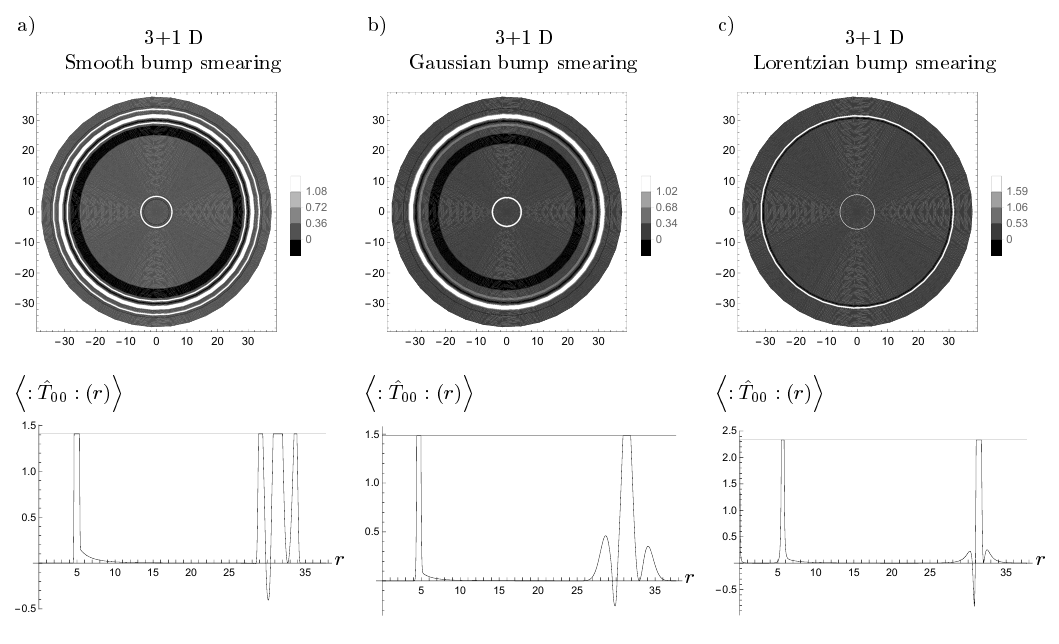}
\caption{Energy density at $x=T+\Delta T$, where Bob's interaction took place at $t=T$. The additional time $\Delta T$ is to allow the ingoing contribution ($r\approx 5$) to move away from the outgoing wavepacket ($r\approx 32$), revealing the negative energy density that propagates radially outward. Unlike the 1+1~D case the ingoing and outgoing wavepackets will respectively increase and decay according to the inverse square law. In the outgoing wavepacket the negative energy well will then decay with time; however, the ratio between the depth of the negative energy well and its neighboring positive energy peak will remain constant. Here a contour plot of the energy density for the $z=0$ surface is shown, accompanied by a radial plot. These three plots (from the left $f,g$ and $h$ smearings) have been truncated so that the divergent nature of the ingoing wave at $r=0$ does not shadow the details of the outgoing wavepacket and its negative energy density.}\label{fig4}
\end{figure*}

As in the case of 1+1~D a series of optimal parameters $(\delta_{\textsc{b}},\sigma_{\textsc{b}},r_{0}^{B},\lambda_{0},\mu_{0})$ were determined for a particular set of $\delta_{\textsc{a}}$ (with $\sigma_{\textsc{a}}=0$ where it applies, by inspection). From the intuition of the 3D wave equation the expectation was that the perturbation created by Alice would propagate outwards and, given the higher dimensionality of the field, would disperse and decay according to the inverse square law. These properties are seen in Fig.~\ref{fig3}. Note that when $t=T$ Alice's outgoing wave overlaps significantly with Bob's smearing and both energy densities before and after Bob's  interaction are shown in Fig.~\ref{fig3}. In this 3+1~D case, since Bob's smearing is a spherically symmetric shell, then the QET protocol creates a shell of negative energy density all around the original position of Alice, which is why in Fig.~\ref{fig3} both the left and right moving waves of Alice have the negative energy wells. Note that as time proceeds following Bob's interaction the amount of negative energy increases as Bob's ingoing positive energy contribution moves away from the negative energy density well. In Fig.~\ref{fig4} we have the radial and contour plots of the energy density for the three spatial smearing functions at a late time.

We see in detail, and for all smearings, in Fig.~\ref{fig4} that for the 3+1~D case the radial outgoing wavepacket, at first sight, has a greater amount of negative energy than that of 1+1~D for comparable widths. Same as discussed before, QET cannot be used to have a negative energy as the leading part of a wavepacket, instead here all the negative energy is inside a shell of positive energy.

\section{Scaling law}\label{sec4}

Numerical inspection quickly points out the existence of a scaling law between the width of the negative energy well with its depth: If we want higher values (in magnitude) of negative energy density, we are bound to have thinner negative energy wavepackets. In this section, we investigate this scaling relation and its consequences with respect to the quantum inequalities.

Let us consider arbitrary compact smearings for Alice and Bob, and scale Alice and Bob's smearing sizes and coupling strengths to see how that affects the energy density distribution they can generate. Our derivation shall be conducted in a general flat $n$-spacetime dimensional  manifold where the generalized $n$ dimensional QET protocol is the natural extension of the 3+1~D QET protocol and is given in full detail in appendix~\ref{seca1}.

Looking  at Eqs. \eqref{eq11} and \eqref{eq50} (or in the appendix, Eqs. \eqref{eq43} and \eqref{eq44} for the $n$-dimensional case) we see that if we want to have an overall positive scaling of the amount of negative energy we should keep $\|\alpha\|$ constant. Not keeping $\|\alpha\|$ constant would yield an exponential decay of the QET term responsible for the negative energy density as $\|\alpha\|$ grows. With that in mind, consider the effect on $\|\alpha\|$ of rescaling the width of Alice's detector smearing function by a factor $\scales^{-1}$:
\begin{align}
\nonumber\|\alpha\|_{\scales}=&\frac{1}{2(2\pi)^{n-1}}\int_{\mathbb{R}^{n-1}}\!\!\!\!\!\!\! \dd^{n-1}\bm{x}\,\int_{\mathbb{R}^{n-1}}\!\!\!\!\!\!\!\dd^{n-1}\,\bm{y}\int_{\mathbb{R}^{n-1}}\!\!\!\!\!\!\! \dd^{n-1}\bm{k}\\
&\qquad\quad\times\lambda(\scales\bm{x})\lambda(\scales\bm{y})\abs{\bm{k}} e^{-2\varepsilon\abs{\bm{k}}-\ii\bm{k}\cdot(\bm{x}-\bm{y})},
\end{align}
and perform a change of variables $\tilde{\bm{x}}=\scales\bm{x}$,  $\tilde{\bm{y}}=\scales\bm{y}$, $\scales\tilde{\bm{k}}=\bm{k}$:
\begin{align}
\nonumber\|\alpha\|_{\scales}=&\frac{1}{2(2\pi)^{n-1}}\int_{\mathbb{R}^{n-1}}\!\!\! \frac{\dd^{n-1}\tilde{\bm{x}}}{\scales^{n-1}}\int_{\mathbb{R}^{n-1}}\!\!\!\frac{\dd^{n-1}\tilde{\bm{y}}}{\scales^{n-1}}\int_{\mathbb{R}^{n-1}}\!\!\!\!\!\!\! \dd^{n-1}\tilde{\bm{k}}\, \scales^{n-1}\\\nonumber
&\qquad\quad\times\lambda(\tilde{\bm{x}})\lambda(\tilde{\bm{y}})\abs{\scales \tilde{\bm{k}}} e^{-2\varepsilon\abs{\scales\tilde{\bm{k}}}-\ii\tilde{\bm{k}}\cdot(\tilde{\bm{x}}-\tilde{\bm{y}})},\\
=&\frac{1}{\scales^{n-2}}\|\alpha\|_{\scales=1},
\end{align}
where one must remember $\varepsilon^{-1}$ is a soft UV cutoff that is ultimately set to infinity.

If we want to keep $\|\alpha\|$ constant, we need to concurrently rescale the coupling strength of Alice's detector as follows: $\lambda(\bm{x})\rightarrow \scales^{\frac{n-2}{2}}\lambda(\scales\bm{x})$. In sum, if we rescale the width of Alice's detector by a factor $\scales^{-1}$, and its coupling strength by a factor $\scales^{\frac{n-2}{2}}$,  $\|\alpha\|$ is left invariant.

We want to address now the following question: Can we make the negative energy distribution arbitrarily large by an appropriate scaling of Bob's detector smearing? if so, does it mean that the support of that distribution would be reduced?

Let us now rescale Bob's smearing function width by the same factor $\scales^{-1}$ and its coupling strength by a factor $\scales^\xi$: \mbox{$\mu(\bm{x})\rightarrow\scales^{\xi}\mu(\scales\bm{x})$}. This rescaling affects the $I_{\mu}^{1}$ terms used to determine the energy density \eqref{eq43} as follows,
\begin{align}
\left(I_{\mu}^{1}(\scales\bm{x},\scales \Delta T)\right)_{\scales}=&\scales^{2\xi}\left(I_{\mu}^{1}(\bm{x},\Delta T)\right)_{\scales =1},
\end{align}
where $\Delta T$ is how much time has transpired since Bob's interaction took place. The fact that we are rescaling the time interval between Alice and Bob's interaction in the same way as their spatial support imposes that Bob's smearing function width also scales by $\scales^{-1}$.

After rescaling (including rescaling time as $t\rightarrow \scales t$) we obtain (see Appendix \ref{seca1}, Eq. \eqref{eq44}), for all times after Bob's interaction ($t>T$), the following rescaled stress-energy density in Eq. \eqref{eq31}:
\begin{widetext}
\begin{align}
\begin{split}
&\bra{\psi(\scales t)}:\hat{T}_{\mu\nu}:(\scales\bm{x})
\ket{\psi(\scales t)}_{\scales}=\Bigg\{\scales^{2 \xi}
\underbrace{\left(I_{\mu}^{1}I_{\nu}^{1}-\eta_{\mu\nu}\frac{I_{\lambda}^{1}I^{1,\lambda}}{2}\right)}_{\text{Bob's positive contribution}}
-\scales^{n}\underbrace{\left(I_{\mu}^{2}I_{\nu}^{2}-\eta_{\mu\nu}\frac{I_{\lambda}^{2}I^{2,\lambda}}{2}\right)}_{\text{Alice's positive contribution}}\\
-&\scales^{\frac{n}{2}+\xi}\braket{A_{0}|\hat{\sigma}_{y}|A_{0}}e^{-2\|\alpha\|}\underbrace{\left(\left(I_{\mu}^{1}I_{\nu}^{3}-\eta_{\mu\nu}\frac{I_{\lambda}^{1}I^{3,\lambda}}{2}\right)+\left(I_{\mu}^{3}I_{\nu}^{1}-\eta_{\mu\nu}\frac{I_{\lambda}^{1}I^{3,\lambda}}{2}\right)\right)}_{\text{QET negative contribution}}\Bigg\}.
\end{split}\label{eq31}
\end{align}
\end{widetext}

In view of this expression, remembering that Bob's coupling strength scales as $\scales^\xi$, we want to choose the scaling power of Bob's coupling strength, $\xi$, so that with changes in scale, the negative energy density is enhanced or at least not overshadowed by the positive energy contributions. Consider the resulting scaled energy density when $\scales\rightarrow \infty$. Bob's positive, Alice's positive and QET's negative contributions will scale as $\scales^{2\xi}$, $\scales^{n}$ and $\scales^{\frac{n}{2}+\xi}$ respectively. In order to avoid the QET's contribution from being overwhelmed by Alice's positive energy contribution we require $\xi\geq n/2$; however, this condition is counterbalanced by the scaling of Bob's contribution. In order for the QET contribution to dominate or at least compete with Bob's contribution we require $\xi\leq n/2$. These two conditions considered together give $\xi=n/2$.  With $\xi=n/2$ we have that the three contributions to the energy density scale as $\scales^{n}$, i.e. scaling does not allow the QET contributions to overpower the positive energy contributions, it only allows for exaggeration of energy densities at the expense of smaller supports for such regions. In other words, we could have arbitrarily large negative energy densities at the expense of thinning out their spatial supports. Enhancing the negative energy density via scaling we found that Bob's coupling strength and smearing function scales as $\mu(\bm{x})\rightarrow \scales^{\frac{n}{2}}\mu(\scales\bm{x})$, while Alice's scales as $\lambda(\bm{x})\rightarrow \scales^{\frac{n-2}{2}}\lambda(\scales\bm{x})$. This results in a scaling relation for the stress-energy tensor,
\begin{align}
\begin{split}
&\bra{\psi(\scales t)}:\hat{T}_{\mu\nu}:(\scales\bm{x})
\ket{\psi(\scales t)}_{\scales}\\
&\qquad = \scales^{n}\bra{\psi(t)}:\hat{T}_{\mu\nu}:(\bm{x})
\ket{\psi(t)}_{\scales=1}.
\end{split}\label{eq35}
\end{align}

To summarize: The scaling relations of the smearing functions that yield an increase of the negative energy density are 
\begin{align}
\lambda(\bm{x})\rightarrow &\scales^{\frac{n-2}{2}}\lambda(\scales\bm{x}),\label{eq37}\\
\mu(\bm{x})\rightarrow & \scales^{\frac{n}{2}}\mu(\scales\bm{x}),\label{eq38}
\end{align}
which lead to the scaling of the stress-energy tensor \eqref{eq35}. Note that, once integrated over space, we have a global linear increase of the total amount of (both) negative (and positive) energy.

Numerical analysis confirms that these scaling relations are indeed the ones that keep optimal QET while scaling up the amount of negative energy in the field. These relations imply that once the optimal QET parameters have been established for a given scenario, we can rescale the setup keeping optimality by appropriately scaling sizes, times and coupling strengths according to \eqref{eq35}, \eqref{eq37} and \eqref{eq38}, which will result in a rescaled negative energy distribution (more peaked and thinner if $\scales>1$).

At the center of the derivation of this scaling law was our desire to create deeper negative energy density wells. For example, let us go back to the case of optimal QET, Fig.~\ref{fig2} and Fig.~\ref{fig4} where we saw the existence of a negative energy well. In the 1+1~D case (Fig.~\ref{fig2}) the scaling relations \eqref{eq35}, \eqref{eq37} and \eqref{eq38} state that given a well of `width' $w$ and `depth' $d$ these can be rescaled to a narrower well of `width' $\frac{w}{\scales}$ and `depth' $\scales^{2} d$, where $\scales>1$. In particular as the well approaches zero-width the total amount of negative energy present in the well diverges as $\scales$. This can be viewed as a scaling relation $\Delta x\propto \frac{1}{\Delta E}$ where $\Delta x$ is the separation between the center of the well and its closest positive pulse and $\Delta E$ is the total negative energy in the well ($\Delta E\sim w\cdot d$). 

Remarkably, this result agrees with the scaling relation presented in the context of the quantum interest conjecture \cite{Ford99}. Note that, unlike in \cite{Ford99}, the present scaling relation was derived from the application of an operational protocol rather than directly from the quantum inequalities. This strongly suggests that optimal QET is very close to saturating the limits imposed by the quantum inequalities, and is therefore a useful tool for operationally exploring these limits.

In the 3+1~D case the scaling relations \eqref{eq37},\eqref{eq38}, when applied to the wells of Fig.~\ref{fig4} show that the well of `width' $w$ and `depth' $d$ can be rescaled to a narrower well of `width' $\frac{w}{\scales}$ and `depth' $\scales^{4}d$. Let us define $\Delta E$ as the total energy density contained in this pulse, which would be roughly $\Delta E \sim \scales^3 w\,d$. Since the scaled pulse has a width $\Delta r=w/\scales$,  we get the scaling relationship $\Delta E\propto \frac{1}{\Delta r^{3}}$, agreeing with the relation given by the quantum interest conjecture \cite{Ford99}, reinforcing the insight discussed above for the 1+1~D case that optimal QET is working close to the limits imposed by the quantum inequalities.

\section{Conclusion}

In this paper we have introduced a QET protocol using the light-matter interaction as a method for generating stress-energy densities capable of violating the classical energy conditions in a controllable way. The violations can be regulated  by adjusting the positions of the qubits in spacetime and their spatial smearing functions, allowing for the possibility of using this protocol as a base for lattices of QET protocols aimed at generating precise stress-energy tensors that violate the classical energy conditions. In particular we focused on the ability of a QET protocol from Alice to a cloud of Bobs operating particle detectors (e.g., atomic probes) in 3+1-dimensional spacetime to generate different negative energy-density distributions.

We have shown that within this protocol there is an inherent scaling relation that allows for the generation of arbitrarily large quantities of negative energy within a well of increasingly narrower support at the expense of neighboring large positive energy wavepackets that avoid a violation of the quantum energy conditions \cite{Ford78,Pfenning98}. We have analyzed the power of these QET protocols to violate classical energy conditions in the light of the quantum interest conjecture \cite{Ford99}. We have shown that the scaling relations derived from qubit QET saturate the scaling laws imposed by the quantum inequalities and the quantum interest conjecture,  indicating that this QET protocol is close to optimal for the generation of negative energy densities.

Our particular application of the QET protocol has produced a situation of a negative energy well pursuing a positive energy wavepacket, as opposed to the protocols involving the dynamical Casimir effect \cite{Ford99} where the negative energy density well precedes a positive energy wavepacket.  The fact that we obtain negative energy surrounded by positive energy distributions stems from causality constraining our placement of Bob to within Alice's lightcone. One future proposal may involve studying a massive scalar field that may overcome this issue by having Alice's main positive energy contribution travel with a group velocity less than light, allowing for Bob to be ahead of it. Since our work is aimed at having Bob as close as possible to Alice's positive wavepacket the exponential decay in correlations in the vacuum state of massive fields should be negligible. Another proposal involves studying a quadratic coupling to the fields in order to produce genuine squeezed states in the field. The quantum energy conditions allow us to anticipate that these states should scale as those presented here (since we obtained optimality); however, with a more favorable coefficient.

Finally, in the light of these results, there are two interesting avenues of research that will be pursued elsewhere.

On the one hand, it would be interesting to study  the fluctuations in the stress-energy density when a QET protocol is used to generate stress-energy distributions that violate energy conditions optimally. Notice that, given the simple coherent-superposition nature of the field state produced by the protocol (See, e.g., \eqref{eqa1}), one would expect these fluctuations to be, in principle, controllable (See, e.g.,\cite{Ford93}). 

On the other hand, after the nature and magnitude of stress-energy fluctuations in this protocol has been established, one could characterize of the behavior of these distributions in the presence of gravity, when the full dynamics enabled by Einstein equations is switched on.


\onecolumngrid

\appendix 
\section{Equivalence of non-local LOQC and distributed LOCC in qubit QET}\label{seca1}

In the text above it was stated that for QET the use of the LOCC variant was equivalent to LOQC. This equivalence allowed us to perform QET with a single, non-locally smeared, Bob under the interpretation that the negative energy density distribution obtained through this process was the same as that obtained under the regular LOCC QET if we considered Bob's smearing as composed of several independent, local agents distributed in space with one qubit detector each \cite{Hotta2008}.

In the following, we will prove that, concerning the expectation value of the renormalized stress-energy tensor (RSET), a LOQC QET scheme---with a single Alice and a single delocalized Bob---is equivalent to a LOCC QET scheme with a single Alice and many localized Bobs properly distributed in space. This will be so when the initial state of Alice's detector is chosen so that the efficiency of QET is maximized (i.e., for the cases considered in this paper, when $\ket{A_0}$ is an eigenstate of $\hat \sigma_y$). This equivalence holds in all dimensions.

\subsection{Stress-energy for LOQC}

Let us begin computing the expectation of the RSET for the LOQC scheme. Consider the state of the field after Alice's interaction and just before Bob's interaction, from \eqref{eq6}:
\begin{align}
\ket{\psi(T^{-})}=&\ket{\bm{\alpha}(T)}\ket{\px_{T}}\braket{\px|A_{0}}+\ket{-\bm{\alpha}(T)}\ket{\mx_{T}}\braket{\mx|A_{0}},
\end{align}
where
\begin{align}
\alpha_{\bm{k}}(T)=&e^{-\ii\left|\bm{k}\right| T e^{-2\varepsilon \left|\bm{k}\right|}}\sqrt{\frac{\left|\bm{k}\right|}{2}}\frac{1}{(2\pi)^{\frac{n-1}{2}}}\int \dd^{n-1}\bm{x}\, \lambda(\bm{x})e^{-\ii\bm{k}\cdot\bm{x}-\varepsilon \left|\bm{k}\right|},
\end{align}
$\ket{\pmx_{T}}$ are the time evolved eigenstates of $\hat\sigma_x$ in time $T$ (under the free detector Hamiltonian $\Omega\hat{\sigma}^{+}\hat{\sigma}^{-}$) and $\varepsilon^{-1}$ is a soft UV cutoff that is ultimately set to infinity. Bob's interaction Hamiltonian takes the form
\begin{align}
\hat{H}_{I,B}=&\delta(t-T) \hat{\sigma}_{z}\int \dd^{n-1}\bm{x}\,\mu(\bm{x})\hat{\phi}(\bm{x}),\label{eqa1p5}
\end{align}
where $\hat{\sigma}_{z}$ acts on Bob's detector, which he received  from Alice via a quantum channel (e.g., quantum teleportation). Following the LOQC protocol, Bob's interaction will result in the state
\begin{align}
\begin{split}
\ket{\psi(T^{+})}
=&e^{-\ii T\Omega}\hat{D}(\bm{\beta})\ket{\bm{\alpha}(T)}\ket{\pz}\braket{\pz|\px}\braket{\px|A_{0}}+\hat{D}(-\bm{\beta})\ket{\bm{\alpha}(T)}\ket{\mz}\braket{\mz|\px}\braket{\px|A_{0}}\\
+&e^{-\ii T\Omega}\hat{D}(\bm{\beta})\ket{\bm{-\alpha}(T)}\ket{\pz}\braket{\pz|\mx}\braket{\mx|A_{0}}+\hat{D}(-\bm{\beta})\ket{\bm{-\alpha}(T)}\ket{\mz}\braket{\mz|\mx}\braket{\mx|A_{0}},
\end{split}\label{eqa1}
\end{align}
where
\begin{align}
\beta_{\bm{k}}=&\frac{-\ii}{\sqrt{2\left|\bm{k}\right|}}\frac{1}{(2\pi)^{\frac{n-1}{2}}}\int \dd^{n-1}\bm{x}\,
\mu(\bm{x})e^{-\ii\bm{k}\cdot\bm{x}-\varepsilon\left|\bm{k}\right|}
\end{align}
and $\Omega$ is the detector energy gap. The resulting energy density is given by the expectation of the RSET:
\begin{align}
\begin{split}
&:\hat{T}_{\mu\nu}:(\bm{x})=\int\frac{\dd^{n-1}\bm{k}\dd^{n-1}\bm{k}'}{(2\pi)^{n-1}\sqrt{4\left|\bm{k}\right|\left|\bm{k}'\right|}}\Bigg\{
\left(k_{\mu}k'_{\nu}
-\eta_{\mu\nu}\frac{k_{\lambda}k'^{\lambda}}{2}
\right)e^{-\varepsilon\left|\bm{k}\right|-\varepsilon\left|\bm{k}'\right|}\bigg[
-e^{\ii(\bm{k}+\bm{k}')\cdot \bm{x}}\hat{a}_{\bm{k}}\hat{a}_{\bm{k}'}+e^{\ii(\bm{k}-\bm{k}')\cdot x}\hat{a}^{\dagger}_{\bm{k}'}\hat{a}_{\bm{k}}\\
+&e^{-\ii(\bm{k}-\bm{k}')\cdot x}\hat{a}_{\bm{k}}^{\dagger}\hat{a}_{\bm{k}'}-e^{-\ii(\bm{k}+\bm{k}')\cdot x}\hat{a}_{\bm{k}}^{\dagger}\hat{a}^{\dagger}_{\bm{k}'}
\bigg]\Bigg\},
\end{split}\\
\begin{split}
&\bra{\psi(T^{+})}:\hat{T}_{\mu\nu}:(x)
\ket{\psi(T^{+})}=\frac{\braket{\pz|\pz}}{2}\frac{1}{4(2\pi)^{2n-2}}\Bigg\{
\left(I_{\mu}^{1}I_{\nu}^{1}-\eta_{\mu\nu}\frac{I_{\lambda}^{1}I^{1,\lambda}}{2}\right)
-\left(I_{\mu}^{2}I_{\nu}^{2}-\eta_{\mu\nu}\frac{I_{\lambda}^{2}I^{2,\lambda}}{2}\right)\\
+&\ii\left(\abs{\braket{\px|A_{0}}}^{2}-\abs{\braket{\mx|A_{0}}}^{2}\right)\left(
\left(I_{\mu}^{1}I_{\nu}^{2}-\eta_{\mu\nu}\frac{I_{\lambda}^{1}I^{2,\lambda}}{2}\right)+
\left(I_{\mu}^{2}I_{\nu}^{1}-\eta_{\mu\nu}\frac{I_{\lambda}^{1}I^{2,\lambda}}{2}\right)\right)\\
+&2\text{Re}\left(\braket{\px|A_{0}}\braket{A_{0}|\mx}\right)e^{-2\|\alpha\|}\left(
-\left(I_{\mu}^{3}I_{\nu}^{3}-\eta_{\mu\nu}\frac{I_{\lambda}^{3}I^{3,\lambda}}{2}\right)+\left(I_{\mu}^{1}I_{\nu}^{1}-\eta_{\mu\nu}\frac{I_{\lambda}^{1}I^{1,\lambda}}{2}\right)\right)\\
-&2\text{Im}\left(\braket{\px|A_{0}}\braket{A_{0}|\mx}\right)e^{-2\|\alpha\|}\left(\left(I_{\mu}^{1}I_{\nu}^{3}-\eta_{\mu\nu}\frac{I_{\lambda}^{1}I^{3,\lambda}}{2}\right)+\left(I_{\mu}^{3}I_{\nu}^{1}-\eta_{\mu\nu}\frac{I_{\lambda}^{1}I^{3,\lambda}}{2}\right)\right)\Bigg\}\\
+&\frac{\braket{\mz|\mz}}{2}\frac{1}{4(2\pi)^{2n-2}}\Bigg\{
\left(I_{\mu}^{1}I_{\nu}^{1}-\eta_{\mu\nu}\frac{I_{\lambda}^{1}I^{1,\lambda}}{2}\right)
-\left(I_{\mu}^{2}I_{\nu}^{2}-\eta_{\mu\nu}\frac{I_{\lambda}^{2}I^{2,\lambda}}{2}\right)\\
-&\ii\left(\abs{\braket{\px|A_{0}}}^{2}-\abs{\braket{\mx|A_{0}}}^{2}\right)\left(
\left(I_{\mu}^{1}I_{\nu}^{2}-\eta_{\mu\nu}\frac{I_{\lambda}^{1}I^{2,\lambda}}{2}\right)+
\left(I_{\mu}^{2}I_{\nu}^{1}-\eta_{\mu\nu}\frac{I_{\lambda}^{1}I^{2,\lambda}}{2}\right)\right)\\
-&2\text{Re}\left(\braket{\px|A_{0}}\braket{A_{0}|\mx}\right)e^{-2\|\alpha\|}\left(
-\left(I_{\mu}^{3}I_{\nu}^{3}-\eta_{\mu\nu}\frac{I_{\lambda}^{3}I^{3,\lambda}}{2}\right)+\left(I_{\mu}^{1}I_{\nu}^{1}-\eta_{\mu\nu}\frac{I_{\lambda}^{1}I^{1,\lambda}}{2}\right)\right)\\
-&2\text{Im}\left(\braket{\px|A_{0}}\braket{A_{0}|\mx}\right)e^{-2\|\alpha\|}\left(\left(I_{\mu}^{1}I_{\nu}^{3}-\eta_{\mu\nu}\frac{I_{\lambda}^{1}I^{3,\lambda}}{2}\right)+\left(I_{\mu}^{3}I_{\nu}^{1}-\eta_{\mu\nu}\frac{I_{\lambda}^{1}I^{3,\lambda}}{2}\right)\right)\Bigg\}.
\end{split}\label{eqa2}
\end{align}
Notice that we have, on purpose, separated this expression in two summands from which the LOCC result will be easily read off later: one where we kept $\braket{\pz|\pz}$ explicitly, and another where we kept $\braket{\mz|\mz}$ explicitly for labeling purposes. The $I^j_\mu$, and $\|\alpha\|$ terms are given by:
\begin{align}
I^{1}_{\mu}(\bm{x})=&\int \dd^{n-1}\bm{y} \dd^{n-1}\bm{k}\,\hat{k}_{\mu}e^{-2\varepsilon\left|\bm{k}\right|}(e^{\ii\bm{k}\cdot(\bm{x}-\bm{y})}+e^{-\ii\bm{k}\cdot(\bm{x}-\bm{y})})\mu(\bm{y}),\\
I^{2}_{\mu}(\bm{x})=&\int \dd^{n-1}\bm{y}\dd^{n-1}\bm{k} \,\hat{k}_{\mu}e^{-2\varepsilon\left|\bm{k}\right|}(e^{\ii\bm{k}\cdot(\bm{x}-\bm{y})-\ii\abs{\bm{k}}T}-e^{-\ii\bm{k}\cdot(\bm{x}-\bm{y})+\ii\abs{\bm{k}}T})\abs{\bm{k}}\lambda(\bm{y}),\\
I^{3}_{\mu}(\bm{x})=&\int \dd^{n-1}\bm{y}\dd^{n-1}\bm{k} \,\hat{k}_{\mu}e^{-2\varepsilon\left|\bm{k}\right|}(e^{\ii\bm{k}\cdot(\bm{x}-\bm{y})-\ii\abs{\bm{k}}T}+e^{-\ii\bm{k}\cdot(\bm{x}-\bm{y})+\ii\abs{\bm{k}}T})\abs{\bm{k}}\lambda(\bm{y}),\\
\|\alpha\|=&\frac{1}{2(2\pi)^{n-1}}\int \dd^{n-1}\bm{x}\dd^{n-1}\bm{y}\,\lambda(\bm{x})\lambda(\bm{y})\int \dd^{n-1}\bm{k}\,\abs{\bm{k}} e^{-2\varepsilon\abs{\bm{k}}-\ii\bm{k}\cdot(\bm{x}-\bm{y})}.
\end{align}
By properly simplifying \eqref{eqa2} the expectation value of the stress-energy tensor assumes the form
\begin{align}
\begin{split}
&\bra{\psi(T^{+})}:\hat{T}_{\mu\nu}:(\bm{x})
\ket{\psi(T^{+})}=\Bigg\{
\left(I_{\mu}^{1}I_{\nu}^{1}-\eta_{\mu\nu}\frac{I_{\lambda}^{1}I^{1,\lambda}}{2}\right)
-\left(I_{\mu}^{2}I_{\nu}^{2}-\eta_{\mu\nu}\frac{I_{\lambda}^{2}I^{2,\lambda}}{2}\right)\\
-&\braket{A_{0}|\hat{\sigma}_{y}|A_{0}}e^{-2\|\alpha\|}\left(\left(I_{\mu}^{1}I_{\nu}^{3}-\eta_{\mu\nu}\frac{I_{\lambda}^{1}I^{3,\lambda}}{2}\right)+\left(I_{\mu}^{3}I_{\nu}^{1}-\eta_{\mu\nu}\frac{I_{\lambda}^{1}I^{3,\lambda}}{2}\right)\right)\Bigg\}.
\end{split}\label{eq43}
\end{align}
For completeness the stress-energy tensor for $t=T+\Delta T$ where $\Delta T>0$ and $T$ is the time of Bob's interaction is given by,
\begin{align}
\begin{split}
&\bra{\psi(t)}:\hat{T}_{\mu\nu}:(\bm{x})
\ket{\psi(t)}=\Bigg\{
\left(I_{\mu}^{1}I_{\nu}^{1}-\eta_{\mu\nu}\frac{I_{\lambda}^{1}I^{1,\lambda}}{2}\right)
-\left(I_{\mu}^{2}I_{\nu}^{2}-\eta_{\mu\nu}\frac{I_{\lambda}^{2}I^{2,\lambda}}{2}\right)\\
-&\braket{A_{0}|\hat{\sigma}_{y}|A_{0}}e^{-2\|\alpha\|}\left(\left(I_{\mu}^{1}I_{\nu}^{3}-\eta_{\mu\nu}\frac{I_{\lambda}^{1}I^{3,\lambda}}{2}\right)+\left(I_{\mu}^{3}I_{\nu}^{1}-\eta_{\mu\nu}\frac{I_{\lambda}^{1}I^{3,\lambda}}{2}\right)\right)\Bigg\}.
\end{split}\label{eq44}
\end{align}
where,
\begin{align}
I^{1}_{\mu}(\bm{x})=&\int \dd^{n-1}\bm{y} \dd^{n-1}\bm{k}\,\hat{k}_{\mu}e^{-2\varepsilon\left|\bm{k}\right|}(e^{\ii\bm{k}\cdot(\bm{x}-\bm{y})-\ii\abs{\bm{k}}\Delta T}+e^{-\ii\bm{k}\cdot(\bm{x}-\bm{y})+\ii\abs{\bm{k}}\Delta T})\mu(\bm{y}),\\
I^{2}_{\mu}(\bm{x})=&\int \dd^{n-1}\bm{y}\dd^{n-1}\bm{k} \,\hat{k}_{\mu}e^{-2\varepsilon\left|\bm{k}\right|}(e^{\ii\bm{k}\cdot(\bm{x}-\bm{y})-\ii\abs{\bm{k}}t}-e^{-\ii\bm{k}\cdot(\bm{x}-\bm{y})+\ii\abs{\bm{k}}t})\abs{\bm{k}}\lambda(\bm{y}),\\
I^{3}_{\mu}(\bm{x})=&\int \dd^{n-1}\bm{y}\dd^{n-1}\bm{k} \,\hat{k}_{\mu}e^{-2\varepsilon\left|\bm{k}\right|}(e^{\ii\bm{k}\cdot(\bm{x}-\bm{y})-\ii\abs{\bm{k}}t}+e^{-\ii\bm{k}\cdot(\bm{x}-\bm{y})+\ii\abs{\bm{k}}t})\abs{\bm{k}}\lambda(\bm{y}).
\end{align}
By inspecting \eqref{eqa2} one can see that the $\braket{\pz|\pz}$ term will equal the $\braket{\mz|\mz}$ term when $\ket{A_{0}}$ is an eigenstate of $\hat{\sigma}_{y}$. We will see below that this is the only condition required for the equivalence in the expectation of the RSET in LOCC and LOQC QET. If we take $\ket{A_{0}}$ to be an eigenvalue of $\hat{\sigma}_{y}$ then \eqref{eqa2} reduces to
\begin{align}
\begin{split}
&\bra{\psi(T^{+})}:\hat{T}_{\mu\nu}:(\bm{x})
\ket{\psi(T^{+})}=\frac{\braket{\pz|\pz}}{2}\frac{1}{4(2\pi)^{2n-2}}\Bigg\{
\left(I_{\mu}^{1}I_{\nu}^{1}-\eta_{\mu\nu}\frac{I_{\lambda}^{1}I^{1,\lambda}}{2}\right)
-\left(I_{\mu}^{2}I_{\nu}^{2}-\eta_{\mu\nu}\frac{I_{\lambda}^{2}I^{2,\lambda}}{2}\right)\\
-&2\text{Im}\left(\braket{\px|A_{0}}\braket{A_{0}|\mx}\right)e^{-2\|\alpha\|}\left(\left(I_{\mu}^{1}I_{\nu}^{3}-\eta_{\mu\nu}\frac{I_{\lambda}^{1}I^{3,\lambda}}{2}\right)+\left(I_{\mu}^{3}I_{\nu}^{1}-\eta_{\mu\nu}\frac{I_{\lambda}^{1}I^{3,\lambda}}{2}\right)\right)\Bigg\}\\
+&\frac{\braket{\mz|\mz}}{2}\frac{1}{4(2\pi)^{2n-2}}\Bigg\{
\left(I_{\mu}^{1}I_{\nu}^{1}-\eta_{\mu\nu}\frac{I_{\lambda}^{1}I^{1,\lambda}}{2}\right)
-\left(I_{\mu}^{2}I_{\nu}^{2}-\eta_{\mu\nu}\frac{I_{\lambda}^{2}I^{2,\lambda}}{2}\right)\\
-&2\text{Im}\left(\braket{\px|A_{0}}\braket{A_{0}|\mx}\right)e^{-2\|\alpha\|}\left(\left(I_{\mu}^{1}I_{\nu}^{3}-\eta_{\mu\nu}\frac{I_{\lambda}^{1}I^{3,\lambda}}{2}\right)+\left(I_{\mu}^{3}I_{\nu}^{1}-\eta_{\mu\nu}\frac{I_{\lambda}^{1}I^{3,\lambda}}{2}\right)\right)\Bigg\}.
\end{split}
\end{align}


\subsection{Stress-energy for LOCC}

Now consider the LOCC case. Following Alice's interaction she performs a measurement of her detector in the $z$ basis and sends her measurement result to Bob. Bob then prepares his detector in the same $z$-eigenstate as the measurement result and couples his detector to the field. In this case, the state prior to Bob's interaction can take one of two forms,
\begin{align}
\ket{\psi(T^{-}),\pz}=&e^{-\ii T\Omega}\ket{\bm{\alpha}(T)}\ket{\pz}\braket{\pz|\px}\braket{\px|A_{0}}+e^{-\ii T\Omega}\ket{-\bm{\alpha}(T)}\ket{\pz}\braket{\pz|\mx}\braket{\mx|A_{0}},\\
\ket{\psi(T^{-}),\mz}=&\ket{\bm{\alpha}(T)}\ket{\mz}\braket{\mz|\px}\braket{\px|A_{0}}+\ket{-\bm{\alpha}(T)}\ket{\mz}\braket{\mz|\mx}\braket{\mx|A_{0}},
\end{align}
depending on whether Alice measured $\pz$ or $\mz$ respectively. These states then evolve with Bob's interaction,
\begin{align}
\begin{split}
\ket{\psi(T^{+}),\pz}
=&e^{-\ii T\Omega}\hat{D}(\bm{\beta})\ket{\bm{\alpha}(T)}\ket{\pz}\braket{\pz|\px}\braket{\px|A_{0}}\\
+&e^{-\ii T\Omega}\hat{D}(\bm{\beta})\ket{-\bm{\alpha}(T)}\ket{\pz}\braket{\pz|\mx}\braket{\mx|A_{0}},
\end{split}\label{eqa4}\\
\begin{split}
\ket{\psi(T^{+}),\mz}=&
\hat{D}(-\bm{\beta})\ket{\bm{\alpha}(T)}\ket{\mz}\braket{\mz|\px}\braket{\px|A_{0}}\\
+&\hat{D}(-\bm{\beta})\ket{-\bm{\alpha}(T)}\ket{\mz}\braket{\mz|\mx}\braket{\mx|A_{0}},
\end{split}\label{eqa5}
\end{align}
 Given the similarities between \eqref{eqa1} and equations \eqref{eqa4}, \eqref{eqa5} one can read off the corresponding expectation values for the stress-energy tensor from \eqref{eqa2},
\begin{align}
\begin{split}
&\bra{\psi(T^{+}),\pz}:\hat{T}_{\mu\nu}:(\bm{x})
\ket{\psi(T^{+}),\pz}=\frac{1}{4(2\pi)^{2n-2}}\Bigg\{
\left(I_{\mu}^{1}I_{\nu}^{1}-\eta_{\mu\nu}\frac{I_{\lambda}^{1}I^{1,\lambda}}{2}\right)
-\left(I_{\mu}^{2}I_{\nu}^{2}-\eta_{\mu\nu}\frac{I_{\lambda}^{2}I^{2,\lambda}}{2}\right)\\
+&\ii\left(\abs{\braket{\px|A_{0}}}^{2}-\abs{\braket{\mx|A_{0}}}^{2}\right)\left(
\left(I_{\mu}^{1}I_{\nu}^{2}-\eta_{\mu\nu}\frac{I_{\lambda}^{1}I^{2,\lambda}}{2}\right)+
\left(I_{\mu}^{2}I_{\nu}^{1}-\eta_{\mu\nu}\frac{I_{\lambda}^{1}I^{2,\lambda}}{2}\right)\right)\\
+&2\text{Re}\left(\braket{\px|A_{0}}\braket{A_{0}|\mx}\right)e^{-2\|\alpha\|}\left(
-\left(I_{\mu}^{3}I_{\nu}^{3}-\eta_{\mu\nu}\frac{I_{\lambda}^{3}I^{3,\lambda}}{2}\right)+\left(I_{\mu}^{1}I_{\nu}^{1}-\eta_{\mu\nu}\frac{I_{\lambda}^{1}I^{1,\lambda}}{2}\right)\right)\\
-&2\text{Im}\left(\braket{\px|A_{0}}\braket{A_{0}|\mx}\right)e^{-2\|\alpha\|}\left(\left(I_{\mu}^{1}I_{\nu}^{3}-\eta_{\mu\nu}\frac{I_{\lambda}^{1}I^{3,\lambda}}{2}\right)+\left(I_{\mu}^{3}I_{\nu}^{1}-\eta_{\mu\nu}\frac{I_{\lambda}^{1}I^{3,\lambda}}{2}\right)\right)\Bigg\},
\end{split}\\
\begin{split}
&\bra{\psi(T^{+}),\mz}:\hat{T}_{\mu\nu}:(x)
\ket{\psi(t),\mz}=
\frac{1}{4(2\pi)^{2n-2}}\Bigg\{
\left(I_{\mu}^{1}I_{\nu}^{1}-\eta_{\mu\nu}\frac{I_{\lambda}^{1}I^{1,\lambda}}{2}\right)
-\left(I_{\mu}^{2}I_{\nu}^{2}-\eta_{\mu\nu}\frac{I_{\lambda}^{2}I^{2,\lambda}}{2}\right)\\
-&\ii\left(\abs{\braket{\px|A_{0}}}^{2}-\abs{\braket{\mx|A_{0}}}^{2}\right)\left(
\left(I_{\mu}^{1}I_{\nu}^{2}-\eta_{\mu\nu}\frac{I_{\lambda}^{1}I^{2,\lambda}}{2}\right)+
\left(I_{\mu}^{2}I_{\nu}^{1}-\eta_{\mu\nu}\frac{I_{\lambda}^{1}I^{2,\lambda}}{2}\right)\right)\\
-&2\text{Re}\left(\braket{\px|A_{0}}\braket{A_{0}|\mx}\right)e^{-2\|\alpha\|}\left(
-\left(I_{\mu}^{3}I_{\nu}^{3}-\eta_{\mu\nu}\frac{I_{\lambda}^{3}I^{3,\lambda}}{2}\right)+\left(I_{\mu}^{1}I_{\nu}^{1}-\eta_{\mu\nu}\frac{I_{\lambda}^{1}I^{1,\lambda}}{2}\right)\right)\\
-&2\text{Im}\left(\braket{\px|A_{0}}\braket{A_{0}|\mx}\right)e^{-2\|\alpha\|}\left(\left(I_{\mu}^{1}I_{\nu}^{3}-\eta_{\mu\nu}\frac{I_{\lambda}^{1}I^{3,\lambda}}{2}\right)+\left(I_{\mu}^{3}I_{\nu}^{1}-\eta_{\mu\nu}\frac{I_{\lambda}^{1}I^{3,\lambda}}{2}\right)\right)\Bigg\},
\end{split}
\end{align}
which in the case that $\ket{A_{0}}$ is an eigenstate of $\hat{\sigma}_{y}$ will equal the LOQC stress-energy tensor,
\begin{align}
\bra{\psi(T^{+})}:\hat{T}_{\mu\nu}:(\bm{x})
\ket{\psi(T^{+})}=\bra{\psi(T^{+}),\pz}:\hat{T}_{\mu\nu}:(\bm{x})
\ket{\psi(T^{+}),\pz}=\bra{\psi(T^{+}),\mz}:\hat{T}_{\mu\nu}:(\bm{x})
\ket{\psi(T^{+}),\mz}.\label{aeq20}
\end{align}

The proof thus far has shown that, provided $\ket{A_{0}}$ is an eigenstate of $\hat{\sigma}_{y}$, then LOCC and LOQC QET give the same stress-energy density. Now we focus on LOCC and consider Bob to be composed of 2 agents, each of which has a detector and an interaction Hamiltonian (c.f. \eqref{eqa1p5})
\begin{align}
\hat{H}_{\textsc{i},\textsc{b}}=&\delta(t-T)\hat{\sigma}_{z,1}\int  \dd^{n-1}\bm{x} \,\mu_{1}(\bm{x})\hat{\phi}(\bm{x})
+\delta(t-T)\hat{\sigma}_{z,2}\int  \dd^{n-1}\bm{x}\, \mu_{2}(\bm{x})\hat{\phi}(\bm{x}).
\end{align}

Following Alice's interaction she performs a measurement of her detector and classically communicates this information to two agents of Bob, who then proceed to prepare their respective detectors. Given that they both receive the same signal their detectors are classically correlated and the resulting state following Bob's interaction is
\begin{align}
\ket{\psi(T^{+}),\pz}
=&e^{-\ii T\Omega}\hat{D}(\bm{\beta}_{1}+\bm{\beta}_{2})\ket{\bm{\alpha}(T)}\ket{\pz_{1},\pz_{2}}\braket{\pz|\px}\braket{\px|A_{0}}+e^{-\ii T\Omega}\hat{D}(\bm{\beta}_{1}+\bm{\beta}_{2})\ket{-\bm{\alpha}(T)}\ket{\pz_{1},\pz_{2}}\braket{\pz|\mx}\braket{\mx|A_{0}},\label{aeq22}\\
\ket{\psi(T^{+}),\mz}=&
\hat{D}(-\bm{\beta}_{1}-\bm{\beta}_{2})\ket{\bm{\alpha}(T)}\ket{\mz_{1},\mz_{2}}\braket{\mz|\px}\braket{\px|A_{0}}+\hat{D}(-\bm{\beta}_{1}-\bm{\beta}_{2})\ket{-\bm{\alpha}(T)}\ket{\mz_{1},\mz_{2}}\braket{\mz|\mx}\braket{\mx|A_{0}},\label{aeq23}
\end{align}
where
\begin{align}
\beta_{\bm{k},1}=&\frac{-\ii}{\sqrt{2\left|\bm{k}\right|}}\frac{1}{(2\pi)^{\frac{n-1}{2}}}\int \dd^{n-1}\bm{x}\,
\mu_{1}(\bm{x})e^{-\ii\bm{k}\cdot\bm{x}-\varepsilon\left|\bm{k}\right|},\\
\beta_{\bm{k},2}=&\frac{-\ii}{\sqrt{2\left|\bm{k}\right|}}\frac{1}{(2\pi)^{\frac{n-1}{2}}}\int \dd^{n-1}\bm{x}\,
\mu_{2}(\bm{x})e^{-\ii\bm{k}\cdot\bm{x}-\varepsilon\left|\bm{k}\right|}.
\end{align}
Having Bob's 2 agents ($\mu_{1}$ and $\mu_{2}$) interacting at the same time simplifies the algebra of the $\hat{D}$ operators. We can define $\bm{\beta}=\bm{\beta}_{1}+\bm{\beta}_{2}$ such that
\begin{align}
\beta_{\bm{k}}=&\frac{-\ii}{\sqrt{2\left|\bm{k}\right|}}\frac{1}{(2\pi)^{\frac{n-1}{2}}}\int \dd^{n-1}\bm{x}
\left(\mu_{1}(\bm{x})+\mu_{2}(\bm{x})\right)e^{-\ii\bm{k}\cdot\bm{x}-\varepsilon\left|\bm{k}\right|}.
\end{align}
Using this new displacement vector we can write equations \eqref{aeq22} and \eqref{aeq23} as
\begin{align}
\ket{\psi(T^{+}),\pz}
=&\left(\hat{D}(\bm{\beta})\ket{\bm{\alpha}(T)}\braket{\pz|\px}\braket{\px|A_{0}}
+\hat{D}(\bm{\beta})\ket{-\bm{\alpha}(T)}\braket{\pz|\mx}\braket{\mx|A_{0}}\right)e^{\ii T\Omega}\ket{\pz_{1},\pz_{2}},\\
\ket{\psi(T^{+}),\mz}=&
\left(\hat{D}(-\bm{\beta})\ket{\bm{\alpha}(T)}\braket{\mz|\px}\braket{\px|A_{0}}
+\hat{D}(-\bm{\beta})\ket{-\bm{\alpha}(T)}\braket{\mz|\mx}\braket{\mx|A_{0}}\right)\ket{\mz_{1},\mz_{2}},
\end{align}
which yield exactly the same field states as a single agent with a non-local smearing $\mu_{1}(\bm{x})+\mu_{2}(\bm{x})$ after performing LOCC QET. $\mu_{1}$ and $\mu_{2}$ can be non-local smearing functions themselves; however, by an induction argument one can see that any non-local smearing function $\mu$ can be decomposed into a sum of local functions, each with a corresponding detector and interaction Hamiltonian.

Hence, in LOCC QET the post-Bob interaction field state from a collection of local agents is identical to the post-Bob interaction field state from a non-local agent provided the sum of the smearings of the local agents coincides with the non-local smearing of the non-local agent, i.e. $\mu(\bm{x})=\sum_{\textsc{i}}\mu_{\textsc{i}}(\bm{x})$.  Then, by virtue of \eqref{aeq20}, we have proven that the stress-energy density in the field after an LOCC protocol with a density $\mu(\bm x)$ of local agents coincides with that resulting from an LOQC protocol involving a single, smeared, non-local Bob with smearing function $\mu(\bm x)$. Therefore we can justify the use of non-local smearings for Bob for the calculations in this paper.



\twocolumngrid

\bibliography{allref}

\end{document}